\documentclass[aip,jcp,reprint,noshowpacs]{revtex4-1}
\usepackage{graphicx}
 \usepackage{amsmath}
 \usepackage{amssymb}
\usepackage{epsfig}
 \usepackage{xcolor}
 \definecolor{Red}{rgb}{0.9,0.0,0.1}
 \usepackage{color, colortbl}

\definecolor{tab1}{cmyk}{0.02,0,0.27,0}
\definecolor{tab2}{cmyk}{0.21,0.04,0,0}

 \usepackage[
         pdftitle={Monte Carlo Simulation of Dense Polymer Melts Using Event
Chain Algorithms},
          pdfauthor={Tobias Kampmann, Horst-Holger Boltz, Jan Kierfeld}
          ]{hyperref}

 \newcommand*{\mat}[1]{\mathbf #1}

 \let \vec \mathbf 

\begin{document}
 \title{Monte Carlo Simulation of Dense Polymer Melts Using Event Chain
   Algorithms}
 \author{Tobias A. Kampmann}
 \email{tobias.kampmann@udo.edu}
 \author{Horst-Holger Boltz}
 \author{Jan Kierfeld}
 \email{jan.kierfeld@tu-dortmund.de}
 \affiliation{Physics Department, TU Dortmund University, 
 44221 Dortmund, Germany}

 \date{\today}

 \begin{abstract}
  We propose an efficient 
Monte Carlo algorithm for the off-lattice simulation of dense
hard sphere 
polymer melts using cluster moves, called event chains, which allow for a
rejection-free treatment of the excluded volume. 
Event chains also allow for an efficient preparation of initial 
configurations in polymer melts.
We parallelize the event chain Monte Carlo algorithm to 
further increase simulation speeds and 
suggest additional local topology-changing moves (``swap'' moves) to 
accelerate equilibration. 
By comparison with other Monte Carlo and molecular dynamics 
simulations,  we verify that the event chain 
 algorithm reproduces the correct equilibrium
 behavior of polymer chains in the melt. 
By comparing  intrapolymer diffusion time scales, 
we show that  event chain  Monte Carlo algorithms  can
achieve simulation
speeds comparable to optimized molecular dynamics simulations.
The event chain Monte Carlo algorithm  
exhibits  Rouse dynamics on short time scales.
In the absence of swap moves, we find reptation dynamics on  
intermediate time scales for long chains.
 \end{abstract}


 \maketitle 

\section{Introduction}

Polymer melts or polymer liquids are  
concentrated solutions of long chain molecules above their
glass or crystallization temperature.
In a dense polymer melt long-range excluded volume interactions become 
 screened and an individual polymer shows ideal behavior.\cite{degennes}
Polymer melts exhibit a
characteristic  and complex dynamical and rheological behavior because of 
entanglement effects, which impede chain diffusion and 
give rise to reptation dynamics 
of polymer chains.\cite{degennes,doiedwards,ferry}
The melt state 
is  also most relevant for processing and manufacturing 
polymer materials.\cite{denn2008}

In this paper we introduce a novel Monte-Carlo (MC)  
  algorithm for the off-lattice 
simulation of a melt of flexible  hard sphere polymer chains,
which are connected by springs or tethers.\cite{Curro1974,
Khalatur1984,Haslam1999,Rosche2000,Karayiannis2008a, Karayiannis2008}
 This  event chain (EC) algorithm allows for a much faster equilibration 
as compared to MC algorithms based on local moves.

The simulation of polymer melts by Molecular Dynamics (MD) or
MC simulations is a  challenging problem,
in particular, for long chains at high density, where polymers in the melt
 exhibit  slow reptation and entanglement dynamics.\cite{doiedwards}
For chain molecules of length $N$, the entanglement
time increases $\propto N^3$, 
which impedes the
equilibration of long chain molecules in a melt if only 
 local self-avoiding 
displacement moves of polymer segments are employed as in a typical
off-lattice MC simulation.
In order to reach equilibrium 
by such local moves, 
the system has to go through slow reptation dynamics on
 time scales between the Rouse and entanglement time.
In MD simulations, such reptation dynamics
has been observed.\cite{Kremer1990,putz2000}
In MC simulations, indications of reptation dynamics
have been observed in lattice models \cite{Kremer1983} or fluctuating
bond lattice models.\cite{Paul1991,kreer2001}
To our knowledge, reptation dynamics has not yet been 
observed in an  off-lattice MC simulation so far, where equilibration 
is more difficult.\cite{GerroffIvoMilchev1993,Binder1997}

The dynamics of MC simulations depends on the 
MC moves that are employed. 
For  local MC moves,
the polymers obey Rouse dynamics on short time scales 
 \cite{Kremer1981,Kremer1983} 
until 
entanglement effects eventually give rise to the crossover to reptation 
dynamics if MC moves obey the self-avoidance 
constraint.\cite{Kremer1983,Paul1991,kreer2001}
This means that the resulting MC dynamics can resemble the actual 
motion  of coarse-grained polymers, although 
the MC dynamics is not explicitly 
based on a realistic  microscopic dynamics.\cite{Binder1997}
Local MC reptation
moves \cite{Wall1975,Kremer1981,Kremer1983,Haslam1999} (slithering snake
moves) are used  to initiate reptation dynamics and obtain faster
equilibration of a polymer melt.  MC simulations have the general advantage
that also non-local or  collective MC moves can be introduced, for
example, chain-topology changing double-bridging 
moves,\cite{karayiannis2002,Karayiannis2008a,Karayiannis2008} which speed
up equilibration (such moves can also be combined with MD simulations 
to equilibrate the system \cite{auhl2003}).
Dynamic properties, however, are no longer realistic 
if such topology-changing moves are employed.
In particular, reptation dynamics will not occur if  chain-topology 
changing moves are employed.

If polymers in a melt are modeled 
as bead-spring models with hard sphere 
beads,\cite{Curro1974,Khalatur1984,Haslam1999,Rosche2000,Karayiannis2008a,Karayiannis2008}
an additional simulation problem  arises, in particular in MC simulations.
At high segment or monomer densities, 
 the mean free path of segments is limited and 
 local MC displacement moves are restricted to very small
step-sizes.\cite{Rosche2000}

For hard sphere systems, non-local  cluster moves represent a 
successful strategy to overcome the problem of slow MC equilibration 
in general by reducing rejection rates in the dense limit. 
In Ref.\ \citenum{krauth2009}, 
the rejection-free event chain algorithm has been proposed, 
which coherently moves large clusters of particles in the form of 
a chain, and a significant speed-up in the sampling of the
hard sphere system has been shown. 
The EC algorithm can be generalized from 
athermal hard sphere systems to spheres with interaction 
potentials.\cite{michel2014,peters2012}
In Ref.\ \citenum{kampmann2015}, we showed that the EC algorithm 
can be used for  simulations of  semiflexible bead-spring polymer  systems. 
In this work, we adapt the EC algorithm for the 
MC simulation of dense polymeric melts consisting of  flexible 
hard sphere polymers, verify 
the algorithm, and benchmark its performance.

The paper is structured as follows: 
in Section \ref{sec:algorithm},  we present our
EC based MC algorithm for hard sphere polymer melts.
In order to further improve performance we also introduce a parallelized 
version of the EC algorithm \cite{kampmann2015} and a version 
employing  local topology-changing ``swap'' moves.
Furthermore, we show that EC moves can also be used to efficiently 
generate  initial polymer configurations for the simulation,
which are already representative of equilibrium configurations. 
In Section \ref{sec:correct}, we verify our algorithm 
by a detailed comparison of equilibrium structural properties, such as
the polymer shape and  the end-to-end distance distribution, 
to simulation results from 
other MD and MC simulation techniques.
Naturally, these results are not novel. Therefore, 
all details of this validation are presented in the Appendix.
Finally, in Section \ref{sec:performance}, 
we benchmark the performance of 
serial and  parallelized EC algorithms with or without swap moves 
against standard local MC schemes and against
state of the art 
MD simulations (using the LAMMPS package \cite{plimpton1995}).
We use time-dependent mean-square displacements (MSDs) of polymer beads to 
monitor inter- and intrapolymer  diffusion and use the intrapolymer 
diffusion to compare the performance of all algorithms
in terms of a   polymer relaxation time. 
The EC algorithm obeys
Rouse dynamics on short time scales. 
Moreover, we show that, in the absence of  topology-changing swap moves
and for long polymer chains,
the EC algorithm exhibits 
 reptation dynamics on intermediate time scales, before a crossover to 
chain diffusion on the longest time scales. 
We end with a conclusion and  outlook.

\section{Event chain  algorithm for polymer melts}
\label{sec:algorithm}

\begin{figure}
 \includegraphics[width=0.95\linewidth]{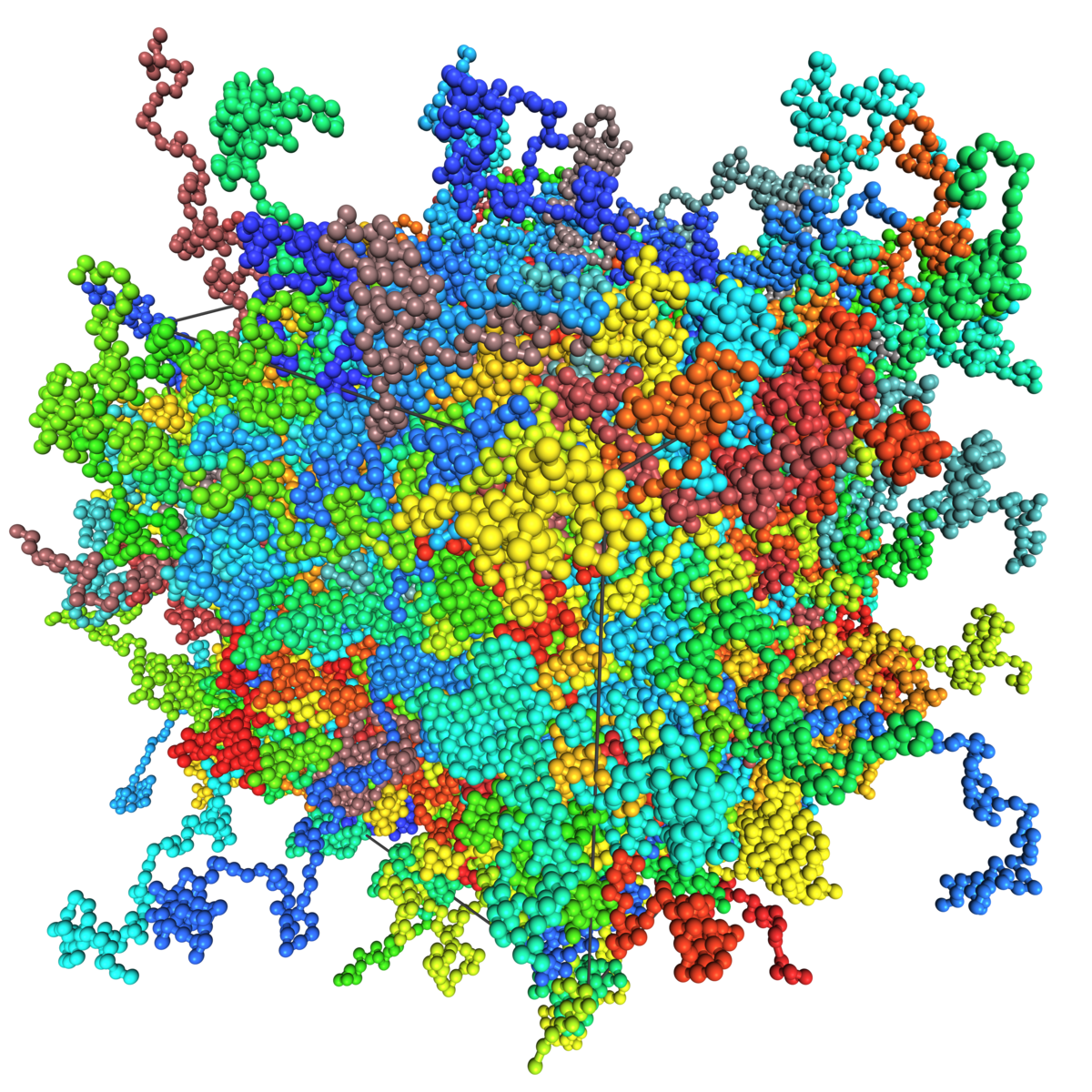}
 \caption{
Simulation snapshot of a polymer melt
 at a volume fraction $\eta \approx 0.54$.  
The color of the beads discriminates the individual
polymers. 
To give better insight into the polymer melt structure, 
we do not wrap polymers periodically although we employ  periodic
 boundary conditions.
}
 \label{fig:melt_snap}
\end{figure}

\begin{figure*}
 \includegraphics[width=0.95\linewidth]{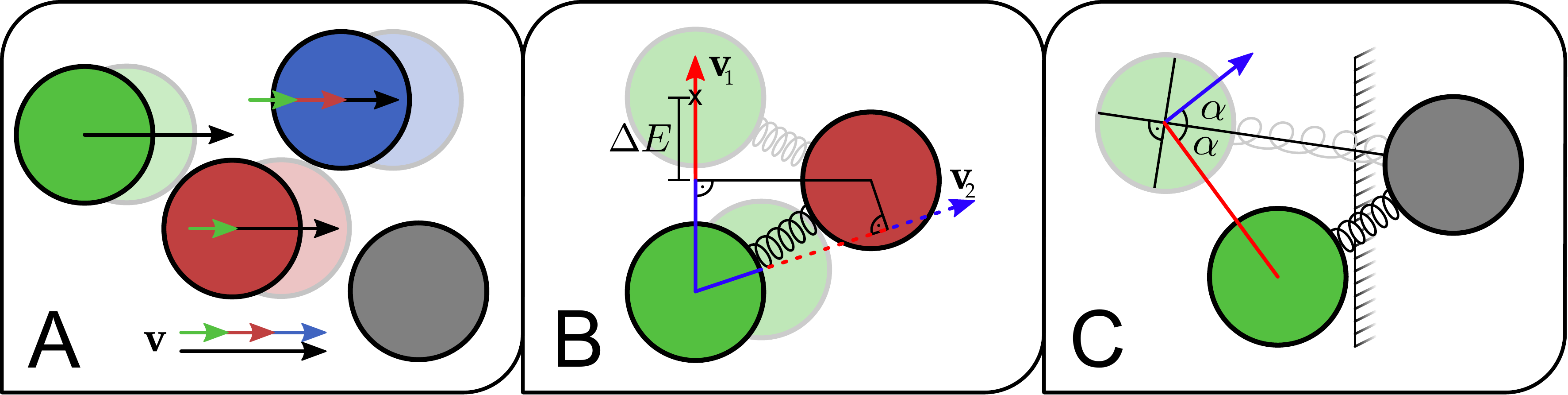}
 \caption{ 
 (a) Construction of an EC cluster in a hard sphere system. 
   The displacement vector $\vec v$ 
  is distributed to all colored beads, 
  which are part of the EC cluster.  
   The final positions are shown in lighter color.
  (b) EC displacements for  harmonic
     bond energies if the green bead is displaced, either in 
   direction $\vec v_1$ without hard sphere collision or in 
   direction $\vec v_2$ where it collides. 
  In both cases, the red bead becomes the next pivot bead.
    For  explanations, see main text. 
  (c)   Reflection of a spring-triggered EC necessary because of the 
    decomposition into parallel simulations cells. 
     The EC cannot be transferred to the gray bead, which 
    is rendered immotile. The pivot bead (green) does
   not change, and the propagation direction 
  is reflected.  For explanations, see main text.
 }           
 \label{fig:EC_algo}
\end{figure*}

A very fundamental model for a self-avoiding flexible polymer is a
bead-spring model, in which all beads interact via an excluded volume
constraint, i.e., 
the polymers consist of hard impenetrable spheres, and the beads in one polymer
are bonded with Hookean springs. 
The spring constant  has to be
sufficiently large as to enforce the impenetrability of polymers and avoid
unphysically large bond stretching.
In summary, we have a hard sphere interaction between all pairs of  beads
\begin{align}
 V(\vec{r},\vec{r}') &= \begin{cases} 
    0 & \lvert \vec{r}-\vec{r}'\rvert > \sigma \\
    \infty & \text{else} \end{cases} \text{,}
\label{eq:Vhardsphere}
\end{align}
with the diameter  $\sigma$ of the hard spheres, and a harmonic stretching
energy which, for a single polymer, can be written as
\begin{align}
 \mathcal{H}_{\text{bonds}} &= \frac{k}{2} \sum_{i=1}^{N-1}(b_i - \sigma)^2
 \text{.}
\label{eq:Hbond}
\end{align}
Here, $k$ is the spring constant, $N-1$ is the number of bonds in a polymer
(containing $N$ beads), 
$b_i$ is the length of the $i$-th bond, and the equilibrium
length of the bonds coincides  with the hard sphere diameter $\sigma$
in our model.
In our simulations, we chose 
bond stiffnesses ($k\sigma^2/k_B T = 30$) such that thermal  bond 
stretching remains weak with
$\langle b_i \rangle \approx 1.1 \sigma$. 
To simulate a polymer melt at a given density $\rho$,
 we generate a system of $M$
polymers in a cube of edge length $L=40\sigma$, see 
Fig.\ \ref{fig:melt_snap}. We
employ periodic boundary conditions in all directions.

Alternatively, we also consider systems of hard sphere polymers
bonded by tethers of maximal length $b_{\rm max} = 1.4\sigma$ 
rather than springs.\cite{Rosche2000}

In  systems of dense hard spheres, standard Metropolis MC schemes
based on local  moves of individual spheres suffer from very slow sampling, as
the move length is limited to roughly the mean free distance between spheres.
This has been
overcome by the introduction of suitable cluster moves, the so-called 
ECs.\cite{krauth2009}
In an earlier work, we extended this approach to parallel computation
and demonstrated how the EC  algorithm can be applied to dense
polymer systems.\cite{kampmann2015} Because the EC algorithm 
moves dense regions of hard spheres or polymer beads coherently, 
it also mimics
the essential features of the actual physical 
dynamics on a coarse time scale such as diffusion of 
polymer bundles.\cite{kampmann2015} If 
the bonded beads in a polymer interact only via pair potentials,
such as in the present example of hard sphere bead-spring polymers, 
 a completely rejection-free simulation solely based on 
EC moves is possible.\cite{peters2012,michel2014}

In the EC algorithm, we first choose a 
 total displacement length $\ell$, which is the same for 
all EC moves. 
For hard spheres, each EC move is constructed according to the following 
rule, which is also illustrated in Fig.\ \ref{fig:EC_algo}(a).
\begin{enumerate}
 \item 
  Select the starting pivot bead for the EC move  and  
  a direction (which we call $\vec v$) randomly. 
 Initially, the remaining displacement is $x_\text{max}=\ell$.
 \item 
  Evaluate the largest possible displacement $x\leq x_\text{max}$
 of the pivot bead in the chosen direction before it touches another bead. 
  Move the pivot bead by $x$.
\item 
  Continue  the EC move at the new pivot bead, which is the 
  hit bead. The remaining 
EC displacement $x_\text{max}$ is decreased by $x$. 
\item 
  Iterate by going back to step 2 until $x_\text{max} = 0$.
\end{enumerate}
Then the next EC move is started.
The relevant computational step in the EC moves
is the evaluation of the admissible
displacement. In a system consisting of unbonded 
  hard spheres, this is the
distance to the bead hit first by the pivot bead while moving in the chosen
direction.

The EC algorithm can be adapted to  spheres with pairwise 
position-dependent 
 interaction potentials.\cite{peters2012,michel2014}
For each move of the pivot bead in an EC chain, an energy 
difference $\Delta E>0$ is  drawn according to the
 Boltzmann distribution.
A displacement of the pivot bead 
that reduces the interaction energy is
accepted 
(as in the standard Metropolis algorithm).
A displacement increasing the  energy  is only
partly executed, up to the point where the 
 energy difference $\Delta E$ that has been drawn is reached
or until the remaining EC displacement has been exhausted.

Now we consider the general situation that the pivot bead has  several 
pairwise interaction energies.
For each interaction partner  $i$, 
the energy difference $\Delta E_i$ then defines a  maximal
displacement of the pivot bead $x_{i}\leq x_\text{max}(\Delta E_i)$. 
The largest possible displacement $x$ of the pivot bead 
is the minimum of all $x_i$, which shall be realized for an interaction
partner $j$, i.e., $x = x_j = \operatorname{min}_i x_i$.
The EC is then  continued at   bead $j$ as next pivot bead.
 For hard sphere interactions, this algorithm reduces to the standard
 EC collision rule.

Fig.\ \ref{fig:EC_algo}(b) shows an example 
for hard sphere polymers bonded by springs.
The attempted EC displacements of the green bead are 
 in one of two classes: (i) the beads
  do not collide along the path (for an EC move 
 in direction $\vec v_1$) or 
 (ii) the beads do
  collide along the path (for an EC move in direction $\vec v_2$). 
  In both cases, the
  energy stored in the bond reduces on the blue part of the trajectory,
  which, therefore, is always admissible, and increases on the red part,
   where the maximal admissible displacement is set by the 
  ``consumable'' energy $\Delta E$ drawn from the Boltzmann distribution. 
Thus, the bond energy is only relevant
    for the maximal displacement if  beads do not collide, because the
     other case is dominated by the hard sphere constraint.
After displacing the green bead, the red bead becomes the pivot bead 
 in both cases (i) and (ii).

We prefer to choose the direction $\vec v$ of ECs randomly, which 
satisfies detailed balance. This can be relaxed, in principle, to 
other choices as discussed in Ref.\ \citenum{krauth2009} for 
hard sphere systems such that global balance is still satisfied. 
One particular simple choice, which can also be applied to the 
 hard sphere polymers, is to start ECs only into  three positive
cartesian directions, which can gain a factor of approximately 2 in 
simulation speed \cite{krauth2009} (essentially by simplifications in 
the collision detection). This simplification
is not efficient, however, if combined   with  the parallelization scheme 
discussed in Sec.\ \ref{sec:Parallelization},
which decomposes the system into simulation cells and 
reflects the ECs on simulation cell boundaries 
rather than rejecting the whole EC move.
For a cell decomposition with rectangular 
boundaries along cartesian directions, as it is usually used,
 EC moves started into cartesian directions  will always 
reflect on themselves.

\subsection{Parallelization}
\label{sec:Parallelization}

We use a parallelized version of this event-chain algorithm and refer to our
earlier work for details  of the parallelization.\cite{kampmann2015}
As discussed there, the parallelization requires a
spatial decomposition of the system
(which is changed in every sweep to ensure ergodicity) into 
 simulation cells.
This  limits the displacement of each sphere to its respective simulation
cell. 
For non-bonded hard spheres this can be treated by reflection of
non-admissible ECs at the cell boundaries.
If a spring-triggered event occurs, where the bonded bead,
which caused the event and would be the next pivot bead, 
is lying outside the current simulation cell, 
we proceed in a very similar manner, i.e.,
by  reflection at the plane normal to the bond of the two participating 
spheres as 
illustrated in Fig.\ \ref{fig:EC_algo}(c):
The gray bead is rendered immotile because of  the currently
   chosen spatial decomposition into parallel simulations cells. 
   Therefore, the EC cannot be transferred to the gray bead
   at the occurrence of a spring ``collision''. Then,
   the pivot bead (green) does
   not change, and the propagation direction 
  is reflected as if there was a wall normal to the bond.

In this work we use a spatial decomposition scheme different from a
checkerboard partition\cite{kampmann2015}: 
we use a rectangular tile-joint
partition, where large tiles are separated by small joints (areas which
contain spheres that cannot move). 
As discussed in Ref.\ \citenum{kampmann2015},
larger cells will lead to a more effective parallelization.

\subsection{Initial configurations}
\label{sec:initial}

The equilibration of polymer melts in simulations 
can  be improved by generating  initial configurations 
that are already representative of equilibrium 
configurations.\cite{auhl2003} Frequently used
strategies consist in a slow compression of an equilibrated 
 dilute solution \cite{Karayiannis2008a,Karayiannis2008} 
or a ``push-off'' procedure, where 
the strongly repulsive steric interaction is switched on only 
 after generating equilibrated  configurations with a soft 
repulsive potential.\cite{auhl2003}
For the hard sphere polymer melt we propose an
EC-based algorithm, which  is 
conceptually similar to the slow push-off
procedure in Ref.\ \citenum{auhl2003} for a Lennard-Jones melt.

\begin{figure}
 \includegraphics[width=0.95\linewidth]{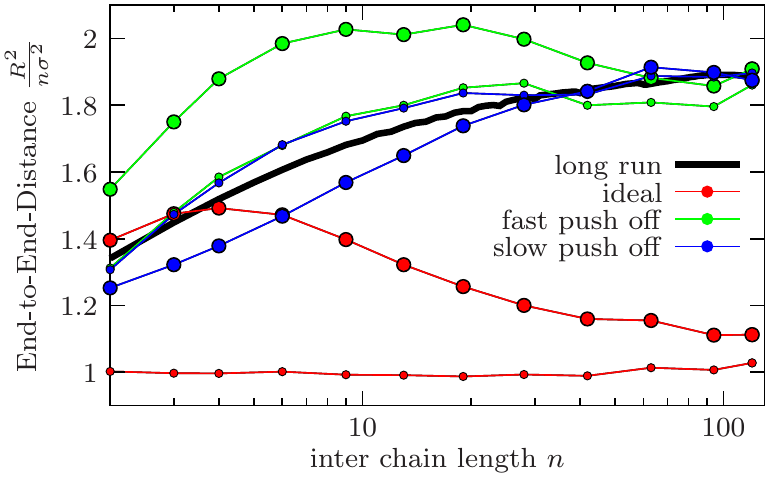}
 \caption{ 
Mean-square internal distances $R^2(n)/n\sigma^2$ between 
  two monomers with a 
 chemical distance $n$ along the chain
 for different
   initial condition generators and  a long run simulation (thick black line). 
  Small points are mean-square internal distances
   directly after setting up the phantom chains,
  corresponding  large points  after increasing the 
   bead size to $\sigma$, either by a ``slow push-off'' 
  (red and blue curves) or by a ``fast push-off'' (green curve). 
  The simulation parameters are $N=121, M=500, L=40\sigma$ 
 (packing fraction $\eta = 0.495$). 
 We used ECs with a total displacement length $\ell = 2 \sigma$ for 
  the push-offs.
}
 \label{fig:rquad_start}
\end{figure}

The flexible polymers in the equilibrated 
melt are ideal  but acquire an effective stiffness.
The effective stiffness is 
 characterized by a finite value of
$\langle \cos\theta\rangle$ ($\theta$ being the bond angle).\cite{auhl2003}
For a long ideal chain of bond length $\sigma$, this results 
in a mean-square end-to-end distance 
$\langle R^2 \rangle (N) = c N\sigma^2$ with  a parameter 
$c\equiv  (1+\langle \cos{\theta}\rangle)/
(1-\langle \cos{\theta}\rangle)>1$,\cite{auhl2003}
 which  depends 
 on the short-range interaction between polymer beads.
From  long-run simulation data, we find $c\approx 1.9$ for a melt of  
long hard sphere chains.

In order to capture the effective stiffness already 
 at the level of the initial
configurations, we set up a system with randomly placed phantom polymers with
vanishing hard sphere diameter and bond length $\sigma$, 
which we grow as non-reversal random walks  
by restricting subsequent (unit) tangents to 
$\vec t_i \cdot \vec t_{i+1} < 
 \cos\left(\theta_\text{max}\right)$.\cite{auhl2003}
For an otherwise uniform 
distribution of bond vectors, this leads to 
  $\langle \cos\theta \rangle = \cos^2(\theta_\text{max}/2)$.
We choose $\theta_\text{max}$ such that 
$\langle R^2 \rangle/N \approx 1.90 \sigma^2$ holds
in accordance with our 
long-run simulation data, see green and blue lines 
 with small symbols  in comparison to 
 black line in Fig.\ \ref{fig:rquad_start}.

We then introduce a finite excluded volume, 
but with a hard sphere diameter that
is only a fraction of the target  diameter $\sigma$. This generates some
``conflicts'', i.e., overlapping spheres.
We  remove these conflicts  
 by repeatedly 
starting EC moves into different directions 
from  the overlapping spheres only,
until the conflicting overlap has been removed.
In these
ECs, we ignore pre-existent overlaps so that the EC will only
be transferred to a bead the current pivot bead is not overlapping with.
This procedure 
corresponds to  locally  ``rattling'' in the hard sphere system until 
enough space has been  created around the overlapping bead 
to insert it. 
Once all conflicts for a given diameter are solved, we increase the diameter
and continue iteratively until the target diameter $\sigma$ is reached.  
The iterative growth of  sphere diameters  (which we call 
 ``slow push-off'' due to conceptual similarity with 
Ref.\ \citenum{auhl2003})  leads to a smaller change 
in the initial distribution of mean-square
internal distances $R^2(n)$ between 
  two monomers with a 
 chemical distance $n$ along the chain (averaged over all chains),
see curves with large symbols in comparison 
to corresponding curves with small symbols in Fig.\ \ref{fig:rquad_start}.
For comparison, we also generate initial configurations  by a 
fast  increase of  $\sigma$ (which we call 
 ``fast push-off'' as in   Ref.\ \citenum{auhl2003}), see green curves in 
Fig.\ \ref{fig:rquad_start}.

Configurations after the push-off
should exhibit internal distances $R^2(n)$
  as close as possible  to the equilibrium result
  as found by a long simulation run, 
   see black line in Fig.\ \ref{fig:rquad_start}. 
The initial configurations generated with slow push-off and the optimal 
value $c \approx 1.9$ (blue lines
in Fig.\ \ref{fig:rquad_start}) are indeed similar to the 
long-run simulation results. 
  The fast push-off configurations (green lines) deviate with  
  a maximum at intermediate $N$, which takes a long time to
 equilibrate due to the slow reptation dynamics.\cite{auhl2003}
Initial configurations generated from   ideally 
flexible phantom chains (red lines)
differ considerably.

Using the slow push-off we can initialize systems at (in principle) any
geometrically possible density  without
resorting to configurations that are far from equilibrium (e.g., placing the
beads on a lattice) in a reasonable amount of time (a couple of minutes
wall time\footnote{We refer to the time needed to perform the simulations as
wall time so as to not be confused with the system time.} for the
system parameters below).
Even for a very dense system with 
$\eta = 0.63995$, $N=120$ and $M =275$, 
we can  generate initial configurations  in $\mathcal O(100h)$ wall time.
For such dense systems, however, these are only 
 {\em valid} configurations, which are  far from equilibrium because  bonds 
are very elongated, and a thorough equilibration is still necessary.

\subsection{Additional Bead Swapping}

\begin{figure}
 \includegraphics[width=0.95\linewidth]{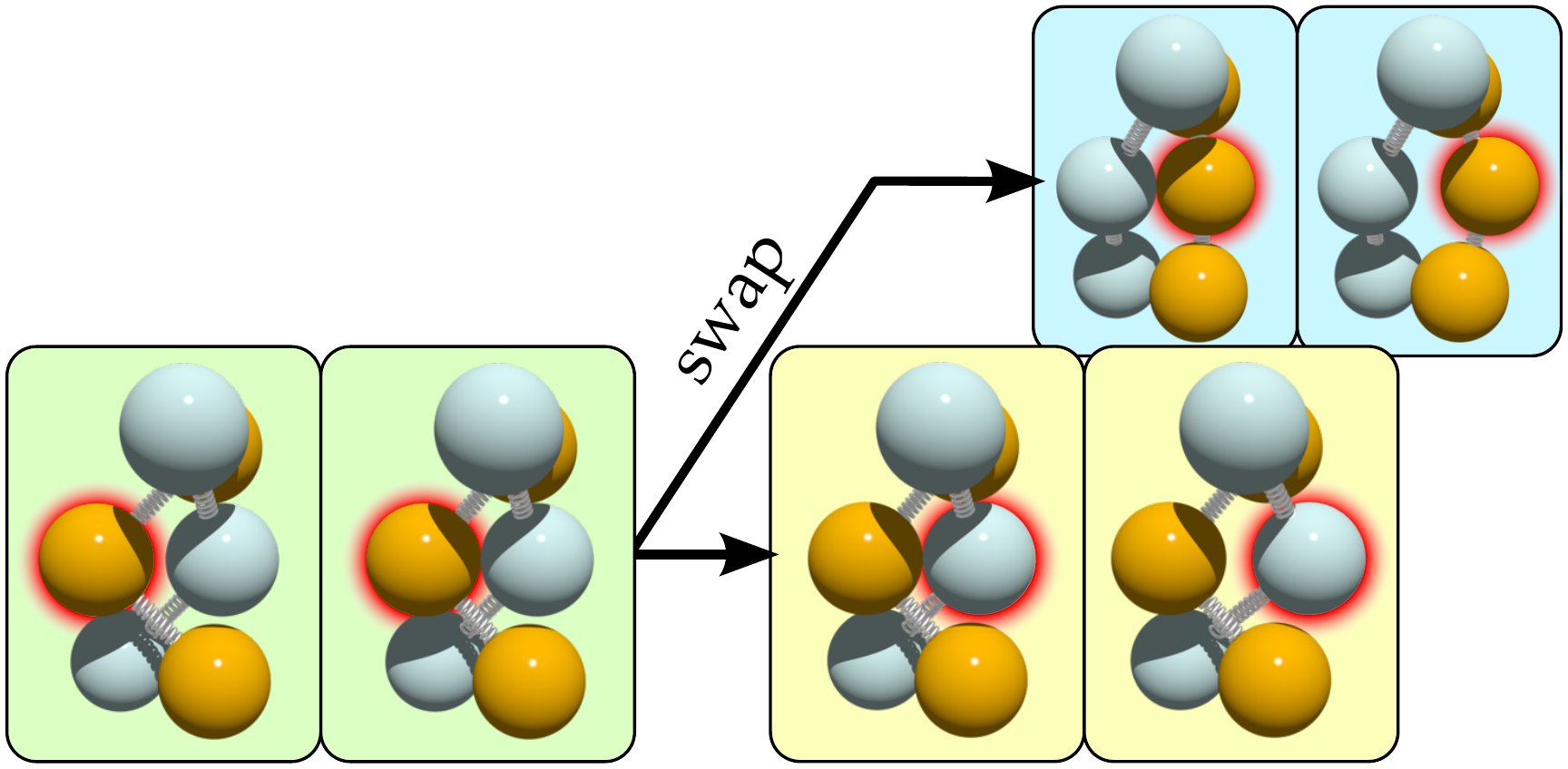}
 \caption{
 From left to right, we show EC moves with/without bead
swapping. 
The currently selected pivot bead is highlighted by a red halo. The
EC  direction is along the vector connecting the middle beads of
the two ``polymers''. The initial EC (first column) is  restricted
by the hard sphere interaction of the middle beads, thus
we use the Metropolis algorithm when the two beads touch 
 (second column), to evaluate whether to swap beads (third column, upper
row) or transfer the EC (third column, lower row). Finally, the
remainder of the displacement is performed (fourth column). 
}
 \label{fig:melt_ref}
\end{figure}

Typical conformations in a dense melt consist of highly entangled
polymers. In the dense limit the ECs become very long, i.e., 
the rather large displacement of an  EC
is distributed  on a lot of very small displacements of many beads
participating in the EC move.
This results in a small collective translation of 
all beads participating in an EC  cluster move
 with only small changes to 
 the  topology of entanglements.

Topology changing MC moves such as the double-bridging 
move \cite{karayiannis2002} can speed up equilibration in 
polymer melts significantly.\cite{auhl2003,Karayiannis2008a,Karayiannis2008}
Here, we improve sampling with EC moves further by introducing an
additional swap MC move, which can locally change the topology 
of entanglements. 
In contrast to the  double-bridging move, which changes bonds, 
  the swap move changes topology by changing bead positions.
 For this purpose, we  modify the EC move so
that the EC does not directly transfer to the next bead upon
hard sphere contact  but, instead, a {\em swap} of the two touching spheres 
is proposed, see  Fig.\ \ref{fig:melt_ref}.
Such an  additional swap move  
allows for a local change of entanglements. 

The EC swap move is  accepted  according to  the standard
Metropolis algorithm. If the swap is rejected, the EC is transferred and
the standard EC algorithm as described above  is recovered.
If it is accepted, the two beads are
exchanged, and the EC continues with  the same pivot bead.
 The example of a swap move in Fig.\ \ref{fig:melt_ref} shows 
a situation where it might be energetically
favorable to swap beads. 
Note that in the absence of bonds all beads become
indistinguishable, and the EC algorithms with and without swapping are
identical up to book-keeping differences.

The swap move is EC-specific: The EC automatically selects
colliding  pairs of beads for swapping; if the swap move is rejected,
the EC move can continue without rejection of the entire EC move. 
Moreover, detailed balance is satisfied, and bead swapping
can be included with very little  computational overhead into 
the EC scheme. 
An analogous swap move
in a standard MC algorithm needs to select 
pairs of beads such that the swap move has a reasonable 
acceptance rate (the particles have to be reasonably close).
Moreover, the selection rule has to satisfy 
detailed balance (for example, simply proposing 
 the nearest neighbor for swapping will lead to a violation 
of detailed balance).
Therefore, there is no straightforward analogue of the EC swap move in 
a standard MC simulation with local moves.

 Since the swap move locally changes topology and 
de-entangles polymers, the dynamics is no 
longer realistic if swap moves are applied.
In particular, reptation dynamics is suppresses by swap moves 
 (see numerical results below).
On the other hand, this is 
 the reason why swap moves can accelerate equilibration of the melt.

\section{Validation}
\label{sec:correct}

 In order to verify our algorithm, 
we address structural equilibrium properties of chains in a polymer melt
by investigating their typical shape as characterized by 
the moment of inertia tensor \cite{karayiannis2009} and 
the distribution of end-to-end distance.\cite{karayiannis2009}
These structural equilibrium quantities provide a detailed comparison 
across polymer melt simulation algorithms. 
The results of a comparison between different MC and MD simulation 
algorithms are shown in the Appendix.
We find  quantitative agreement between the EC algorithm and 
standard MC and MD algorithms,  and agreement with 
 previous MC simulation results and theoretical predictions, where available.

\section{Performance and Dynamics}
\label{sec:performance}

\begin{figure*}
 \includegraphics[width=0.95\linewidth]{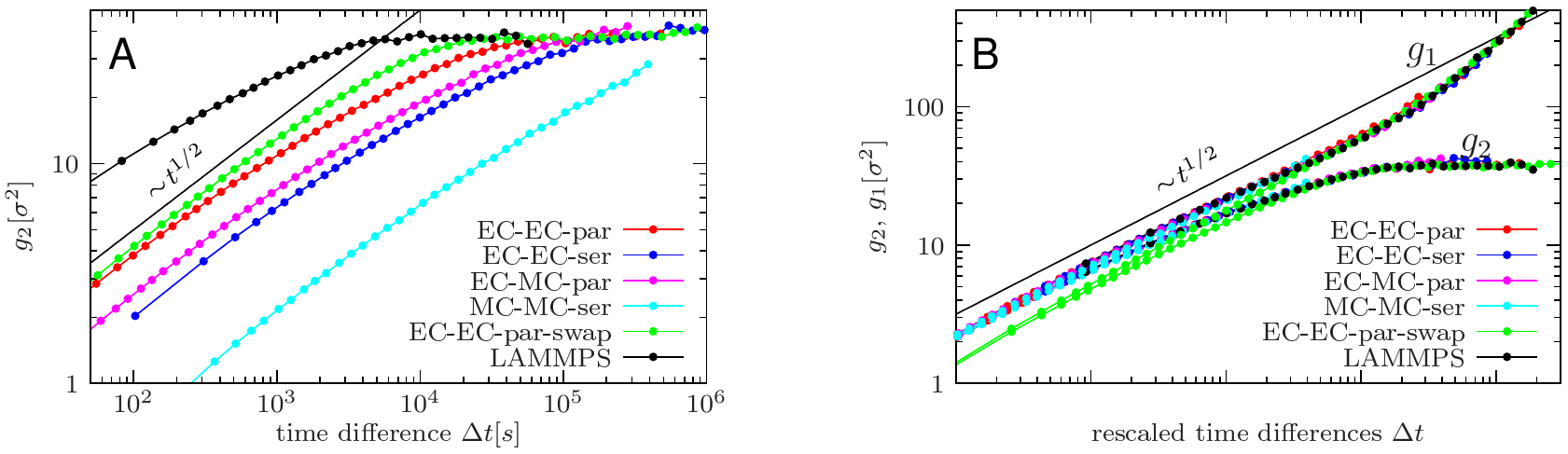}  
 \caption{
 (a) 
  Relative  mean-square displacement $g_2(\Delta t)$ 
  (log-log plot)   for
   different algorithms against wall time (in s). 
  $g_2$ approaches a plateau value, which defines the  polymer relaxation time 
scale, which serves as a measure for  algorithm speed.  
 (b) The  mean-square displacements  $g_1(\Delta t)$ 
  and    $g_2(\Delta t)$  (log-log plots) for
   different algorithms  with time 
  rescaled to collapse all curves. Collapse is achieved for  
  all algorithms except the one using swap moves. 
  All simulations were performed for System I. 
}
 \label{fig:g23_wall}
\end{figure*}

For the comparison of the performance of different algorithms,   
we distinguish  algorithms by  whether they use
  the EC or standard Metropolis algorithm for (i) the hard sphere
 interactions and/or (ii) the bond spring interactions 
   (``EC'' for event chain, ``MC'' for standard Metropolis) and (iii) if
  the algorithm is executed parallelly (par) or serially (ser) and 
  (iv) if the swap  move is used (swap).    
   Accordingly, we introduce a naming scheme for algorithms where, 
   for instance, ``EC-MC-par'' refers to a parallelized
  simulation, where hard sphere interactions  are handled by the EC, 
   springs handled by standard Metropolis algorithm, and the swap move 
 is not used.

 We  compare five different algorithms,  namely EC-EC-par-swap,
 EC-EC-par, EC-EC-ser, EC-MC-par, and MC-MC-ser. This allows us to
  analyze the parallelization performance gains by  comparing
 EC-EC-par/ser and check if we achieve the 
 theoretical speed-up  factor given by the number of  processor cores.
  We do not parallelize the standard MC algorithm, because
 it was shown previously that strong scaling is
 achievable.\cite{anderson2013}
 The comparison of
   MC-MC-ser/EC-EC-ser gives the algorithmic speed-up by using 
  the event chain  algorithm.
 The comparison EC-EC/MC-par demonstrates the advantage of using the
 event chain on the pair potential, i.e., the bonds.

Additionally, we compare our results to those from MD simulations
performed using the highly optimized LAMMPS package.\cite{plimpton1995}
As hard spheres 
cannot be used in a force-based MD simulation, we compare with 
 beads that are interacting via the
repulsive part of standard Lennard-Jones potentials, whereas the bonds remain
Hookean springs. The identification of the effective hard sphere radius 
of such soft Lennard-Jones spheres has been subject of prior 
research.\cite{andersen1971}
Our results  show that  identifying the Lennard-Jones length scale 
$\sigma_{\text{LJ}}$ 
(defined by the zero of the full Lennard-Jones potential,
 $V_{\text{LJ}}(\sigma_{\text{LJ}})=V_{\text{LJ}}(\infty)=0$)
 with  the hard sphere diameter $\sigma$ 
suffices for the purposes of this work. 
This comes with the advantage that we can use the same
initial configurations (generated by our EC-based procedure)
for the MD and MC evolutions.

\subsection{Diffusional dynamics and algorithm speed}

We compare the speed of different algorithms in terms of  wall time. 
 Since the simulations ran on different Central Processing Units 
(CPUs),
 all wall times were calibrated with short run simulations on the same 
 workstation
 with four CPUs for comparable results.

We choose three different systems to investigate the influence of the occupied
volume fraction $\eta$ and chain length $N$ on  algorithm performance 
(we use the same systems
 for the validation of equilibrium properties in the Appendix):
\begin{enumerate}
\item 
System I: $M=400$, $N=120$ and $\eta = 0.390$; 
\item 
System II: $M=500$, $N=120$ and $\eta = 0.490$;
\item
System III: $M=250$, $N=240$ and $\eta = 0.490$.
\end{enumerate}
This means the volume fraction $\eta$ increases  from  System I to 
System II, whereas the polymer length $N$ increases when going from 
  System II to  System III.

In the following, we will compare the performance of these 
 algorithms by the inter- and intrapolymer
diffusional behavior of polymer chains using 
 time-dependent MSDs of polymer beads. 
 For a chain with
bead positions $\vec{r}_i$ ($i=1,\ldots,N$) and center of mass $\vec{R} =
\frac{1}{N} \sum_{i=1}^N \vec{r}_i$, we measure the  
MSD functions,\cite{putz2000,kreer2001}
\begin{align}
 g_{1}(\Delta t) &= \langle \left[ \vec{r}_{N/2}(t+\Delta t) 
      - \vec{r}_{N/2} (t)  \right]^2 \rangle_t, \\
 g_{2}(\Delta t) &= 
  \langle \left[ (\vec{r}_{N/2}(t+\Delta t) - \vec{R}(t+\Delta t))
    \right. \nonumber\\ 
   & \left.  ~~~~ - (\vec{r}_{N/2} (t) - \vec{R}(t)) \right]^2 \rangle_t.
\end{align}
$g_{1}$ 
describes the diffusion of the middle bead  including contributions 
from inter- and intrapolymer diffusion and 
 $g_{2}$ the intrapolymer diffusion  of the middle bead 
 relative to the center of mass of the polymer.
For both quantities, the average  $\langle \ldots \rangle_t$ 
is an ensemble average and an average over  time.

In a polymer melt, 
the time evolution is governed by a 
sequence of crossovers,\cite{degennes,doiedwards}
\begin{align}
 g_1(t) &\sim \begin{cases}
		  t^{1/2} \quad \text{ for } t < \tau_e \\
		  t^{1/4} \quad \text{ for }  \tau_e < t < \tau_R \\
		  t^{1/2} \quad \text{ for }  \tau_R < t < \tau_d \\
		  t\phantom{^{1/1}}  \quad  \text{ for }  \tau_d < t \\
	      \end{cases}, 
\label{eq:g1}
\end{align}
with three different crossover time scales: the 
 entanglement time scale  $\tau_e$, the Rouse time scale $\tau_R$, and 
the disentanglement time scale $\tau_d$.\cite{putz2000}
For all times scale $t>\tau_e$, reptation slows down the 
diffusional dynamics. 
The relative 
MSD $g_2$ exhibits the same regimes as $g_1$
but is insensitive to center of mass diffusion. 
For $t> \tau_d$, it approaches  a plateau value given by 
 the radius of gyration 
$R_g^2(N) = N^{-1}\sum_{i=0}^{N} \langle(\vec{r}_i - \vec{R})^2\rangle$.

Any simulation dynamics achieving equilibration of intrapolymer 
 modes, will reach  the   plateau in  the relative 
MSD $g_2(t)$, beyond which intrapolymer 
 fluctuations are equilibrated. 
We use the relaxation time to reach the plateau
as a measure of simulation speed because it characterizes the 
equilibration
performance of an algorithm on the scale of whole polymer chains. 
If the algorithm correctly describes the polymer melt 
dynamics on long time scales and 
exhibits Rouse, reptation, and chain diffusion dynamics 
as in Eq.\ (\ref{eq:g1}),  this relaxation 
time will coincide with the polymer disentanglement time $\tau_d$. 
In  Fig.\ \ref{fig:g23_wall}(a), we  show the wall time  evolution of
 $g_2(t)$ for different algorithms. 
Both  MD  (LAMMPS) and  
local MC dynamics 
follow Rouse dynamics with a $t^{1/2}$-behavior for short 
times.\cite{Kremer1983,Kremer1990}
 Remarkably, we 
find such  Rouse dynamics also for the  cluster EC algorithm,
even in the presence of  of swap moves. All algorithms 
approach  a  plateau in  the relative 
MSD $g_2(t)$.

 This also allows us to easily compare the performance 
of the algorithms and to determine a speed-up factor for each algorithm
 by  rescaling  time, i.e., 
shifting the double logarithmic 
curves such that the curves $g_2(t)$ coincide
for long time scales close to the plateau. 
As a result, the  
single polymer relaxation time, which is identical to the
disentanglement times $\tau_d$ if the algorithm 
exhibits all characteristic regimes of polymer melt dynamics, 
 should be identical after rescaling. 
The resulting speed-up factors with respect to the 
standard local 
Metropolis algorithm  MC-MC-ser are shown in Table \ref{tab:effvgl}.

If we use these speed-up factors for a 
linear  rescaling of the time, the data for {\em both} MSD
 functions $g_2(t)$ and $g_1(t)$ and from all algorithms
collapse onto two ``master curves'' as shown in 
Fig.\ \ref{fig:g23_wall}(b). The exception is the 
EC algorithm employing topology-changing swap moves.
Because this collapse includes the MD algorithm, 
this  provides evidence 
that both local MC dynamics and the cluster  EC dynamics  
(in the absence of swap moves)
  evolve the system in a
way that allows for an  identification of  ``Monte Carlo time'' 
(i.e., number of moves) with physical time.

\begin{table}
\begin{ruledtabular}
\begin{tabular}{lrrr} 
algorithm & System I & System II & System III \\
\colrule
 MC-MC-ser & $1$ & $1$& $1$\\
EC-EC-ser & $9$ & $7$ & $ 7$\\
EC-MC-par & $14$ & $17$ & $17$ \\ 
EC-EC-par & $31$ & $25$ & $28$ \\
EC-EC-swap-par & $ 230$ & $115$ & $423$ \\
LAMMPS (par) & $330$ & $625$& $770$ \\ 
\end{tabular}
\end{ruledtabular}
\caption{
  Comparison of relative speed-up compared to the standard local 
  Metropolis algorithm  MC-MC-ser for different system parameters 
  and algorithms. 
 Both parallel EC and LAMMPS simulations were performed using four cores. 
}
\label{tab:effvgl}
\end{table}

For the comparison in Table \ref{tab:effvgl}, 
we did not explicitly optimize the free simulation parameters like the
total displacement length $\ell$ for an EC 
 or the number of started ECs per sweep  in a parallelized
simulation (see Ref.\ \citenum{kampmann2015} for a detailed discussion).  
 Nevertheless, it is obvious that all algorithms clearly outperform
the standard MC algorithm.
Without parallelization,  the EC-EC-ser algorithm achieves 
speed-up factors up to 10 compared to the standard 
MC algorithm (MC-MC-ser).
The parallelization gives an additional  speed-up  factor of
  $3.5 \ldots 3.9$ close to the theoretical limit of $4$
given by the number of cores we used for the parallel simulation.
 We note that also standard MC  algorithms could be 
parallelized such that this additional parallelization 
speed-up factor is not specific to the EC algorithms.

Despite these speed-up factors for the EC algorithm, 
the LAMMPS MD simulation  is still the fastest algorithm. 
For the comparison in in Table \ref{tab:effvgl},
we used a parallel version of LAMMPS running on four cores.
We note that
 LAMMPS is under development since the mid 1990s\cite{plimpton1995}
whereas our EC algorithm implementation, 
while adhering to general good practice rules for
scientific computation, should still have  room for optimization.
In view of these preliminaries, the performance difference between the MD
LAMMPS simulation and our fastest EC variant including swap 
moves seems very promising.

Table \ref{tab:effvgl} also shows that 
the EC-MC
  algorithm gains some efficiency with  respect to  EC-EC algorithms 
with increasing
  $\eta$. In  such dense systems, the springs are 
  compressed to a value close to their rest length $\sigma$. 
 Therefore, the
  rejection rate caused by the spring energy is rather low, such that the
  gain from the 
   additional computational effort in the rejection-free treatment
 of springs   is small  in denser systems. 
  Since the disentanglement time $\tau_d\sim N^3$ is
  strongly influenced by the chain length $N$, the efficiency of the swap
  algorithm increases with longer chains.
MD performance does not decrease with density, whereas 
 local MC and also EC performance depends on the displacement 
 length $\ell$, which decreases with density. 
This  explains the  performance
differences if the  density $\eta$ is increased.

The speed-up factors in Table \ref{tab:effvgl} 
characterize algorithm equilibration times based on the 
polymer disentanglement time $\tau_d$. Alternatively, 
the autocorrelation time of the  end-to-end vector
can  been used to characterize equilibration 
times.\cite{GerroffIvoMilchev1993,karayiannis2002,Karayiannis2008}
According to Ref.\ \citenum{Wittmer2007}, these equilibration 
time scales 
are comparable for moves not changing the chain topology; the 
disentanglement time $\tau_d$ is to be preferred if topology-changing 
moves are employed that cannot relax density fluctuations
(e.g., double bridging moves or our swap move).

\subsection{Reptation dynamics}

\begin{figure}
 \includegraphics[width=0.95\linewidth]{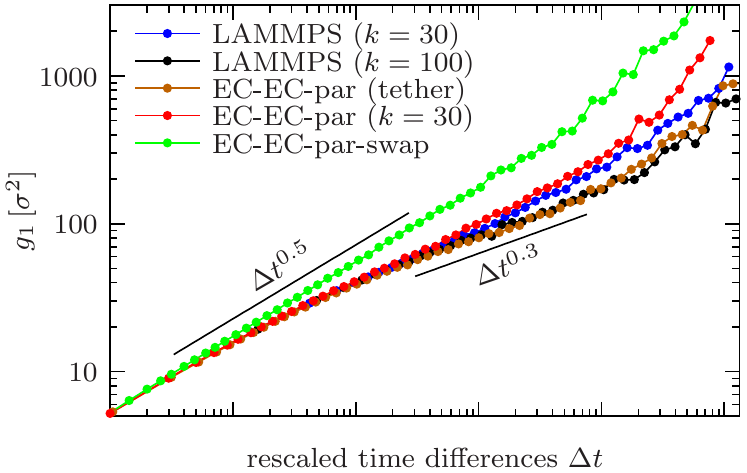}
 \caption{ 
  Mean-square displacement  $g_1(\Delta t)$ (log-log plot) 
     for different algorithms  with time 
  rescaled to collapse all curves for $M=20$ polymers with length $N=500$
  (values for the spring constant $k$ in units of $k_B T / \sigma^2$). 
   }
 \label{fig:g1_longpol}
\end{figure}

The good collapse onto master curves in Fig.\ \ref{fig:g23_wall}(b)
suggests that we can observe the same regimes of polymer melt dynamics
in the EC simulation as in a MD simulation, at least 
if no swap moves are employed. Therefore, we investigated whether 
also a regime of reptation dynamics is observable
with the EC algorithm.

The reptation regime for $\tau_e < t < \tau_R$ is rather hard to observe 
in simulations of 
shorter chains, and one expects a slightly increased exponent $g_1(t) \sim
t^{x}$ with $ 0.25 \leq x < 0.4$.\cite{ebert1997}  
For the chain lengths $N=120$ used
in Fig.\ \ref{fig:g23_wall}(b), the intermediate reptation 
regime  is not clearly visible. 
In lattice MC simulations, evidence for an intermediate reptation-like
regime with a considerably
slower increase than $t^{1/2}$ has only been found 
in  melts with  long chains of length
of $N=512$.\cite{kreer2001}

Therefore, we also simulated a 
 smaller system with less ($M=20$)  but longer polymers ($N=500$)
 at $\eta = 0.298$
and measured $g_1(t)$  for EC-EC algorithms in comparison with MD 
simulations (using LAMMPS), see Fig.\ \ref{fig:g1_longpol}.
For these longer chains, the MD simulations  show  a much more pronounced 
intermediate regime of slowed down dynamics.
We find that this regime is  increasing for stiffer 
polymer springs  
 ($k=100 k_B T / \sigma^2$ as compared to $k=30 k_B T / \sigma^2$
in Fig.\ \ref{fig:g1_longpol}):
with a small probability, chains can still cross via  thermally 
activated bond stretching, which becomes less probable 
for stiffer springs. 
For stiff springs ($k=100 k_B T / \sigma^2$), 
the MD simulations exhibit a reptation regime with a 
time-dependence $g_1(t) \propto t^{0.3}$ close to $t^{1/4}$ and 
 in accordance with theoretical predictions and simulations in 
  Ref.\ \citenum{ebert1997}.

The parallelized EC-EC simulations 
show exactly the same dynamical regimes in the MSD function 
 $g_1(t)$ as the MD simulations, 
see Fig.\ \ref{fig:g1_longpol}.
In each EC move, all beads  that would
collide successively during a short time interval 
in  a MD simulation are displaced  
at once. This gives rise to a MC dynamics 
which is effectively very similar to the realistic MD dynamics. 
For the EC-EC simulations, we considered 
a spring constant  $k=30 k_B T / \sigma^2$ and, instead of stiff springs, 
 hard sphere polymers
bonded by tethers of maximal length $b_{\rm max} = 1.4\sigma$.
In the tethered system, thermally activated bond crossing 
is absent similarly to a system with very stiff springs. 
Very similar to the stiff spring MD simulation, the tethered EC-EC 
simulation shows  evidence of an intermediate  reptation regime 
with  $g_1(t) \propto t^{0.3}$. To our knowledge, this is 
the first off-lattice  MC simulation, where clear indications of 
 reptation dynamics  could be observed.

Fig.\ \ref{fig:g1_longpol} also shows, that 
the reptation regime is absent as soon as we employ additional 
disentangling swap moves in accordance with our expectation.
Swap moves can thus be used to accelerate equilibration by 
effectively ``switching off'' the slow reptation dynamics.

\section{Conclusion}

We introduced novel efficient off-lattice
MC  algorithms for the simulation of 
dense polymer melts of hard sphere polymers, 
which are based on event chain  cluster moves previously known 
for hard sphere systems, see Fig.\ \ref{fig:EC_algo}.
These EC cluster moves  allow for a
rejection-free treatment of the excluded volume interaction 
in the polymer melt. We generalize the algorithm to also handle 
the spring interactions in polymer bonds rejection-free. 

In addition, we introduce  an efficient procedure to 
generate initial configurations, which are representative 
of typical  equilibrated configurations in polymer melts. 
Using EC ``rattling'',
we can generate initial configurations up to very high packing fractions
(up to $\eta = 0.63995$). 

We parallelize the event chain Monte Carlo algorithm and 
suggest additional local topology-changing swap moves,
see Fig.\  \ref{fig:melt_ref},  to further increase 
simulation speeds in melts.

We validated the EC algorithm by  
 comparing equilibrium structural properties.
In the Appendix, we show results for 
the polymer shape (Fig.\ \ref{fig:traeg}) 
and  the end-to-end distance distribution
(Fig.\ \ref{fig:rquad_dist}),
which are in quantitative agreement with 
other MD and MC simulation techniques.

We assessed the performance of the EC algorithm by measuring its 
equilibration speed using the relative  MSD function $g_2(t)$ of 
a polymer bead in the middle of a polymer with respect to the 
polymer center of mass, see Fig.\ \ref{fig:g23_wall}. 
This  allows us to define 
a polymer relaxation time, which is specific to the algorithm 
and represents a measure for its  equilibration speed. 
We find that the combination of EC moves and parallelization
can increase  MC  simulation speeds by factors up to 30.
If also swap moves are employed, MC simulation speeds become comparable
to optimized MD simulations
that we performed with the LAMMPS package for comparison.

Without swap moves, the dynamics of the EC algorithm appears 
to be very similar to MD dynamics. A simple rescaling 
of simulation times can collapse MD and EC simulation dynamics,
 see Figs.\  \ref{fig:g23_wall} and \ref{fig:g1_longpol}.
The collective dynamics generated by the EC moves, which essentially
displace all beads coherently that collide successively 
in a short time interval in a MD simulation, 
appears to be very similar to the MD 
collision dynamics.

Accordingly, 
in the absence of swap moves, the EC algorithm  exhibits all 
dynamical regimes expected for polymer melts, i.e., 
 Rouse, reptation, and chain diffusion dynamics.
 In particular, we can identify  an intermediate 
 reptation regime with a MSD function
 $g_1(t) \propto t^{0.3}$ close to $t^{1/4}$
in simulations
of a system with long chains ($N=500$), see Fig.\ \ref{fig:g1_longpol}.
 To our knowledge, this is  
the first off-lattice  MC simulation, where  reptation dynamics 
could be observed. 
If topology-changing swap moves are used, which disentangle 
polymer chain, reptation dynamics is absent in the EC algorithms.

Although we only presented results for the most simple case of a
melt of flexible polymers with no interpolymer interaction other than excluded
volume, the added value of EC algorithms should persist in more
complex systems. 
For (bond) interactions that are not pair interactions, e.g.,
bending energies, rejection-free sampling in the way presented here is not
possible. We have already shown in a previous work,\cite{kampmann2015}
however, 
that such bending energies can still be
treated by proposing moves that are compliant with the hard sphere constraint
by using ECs and than accepting (or declining) this move according to
the standard Metropolis algorithm.

\vspace{0.5cm}

\begin{acknowledgments}

We acknowledge financial support by  the Deutsche Forschungsgemeinschaft
(No.\ KI 662/2-1).

\end{acknowledgments}

\vspace{0.5cm}

\bibliography{meltropolis}

\appendix
\section{Validation}
\label{app:Validation}

\subsection{Moment of inertia tensor}

\begin{figure}
 \includegraphics[width=0.9\linewidth]{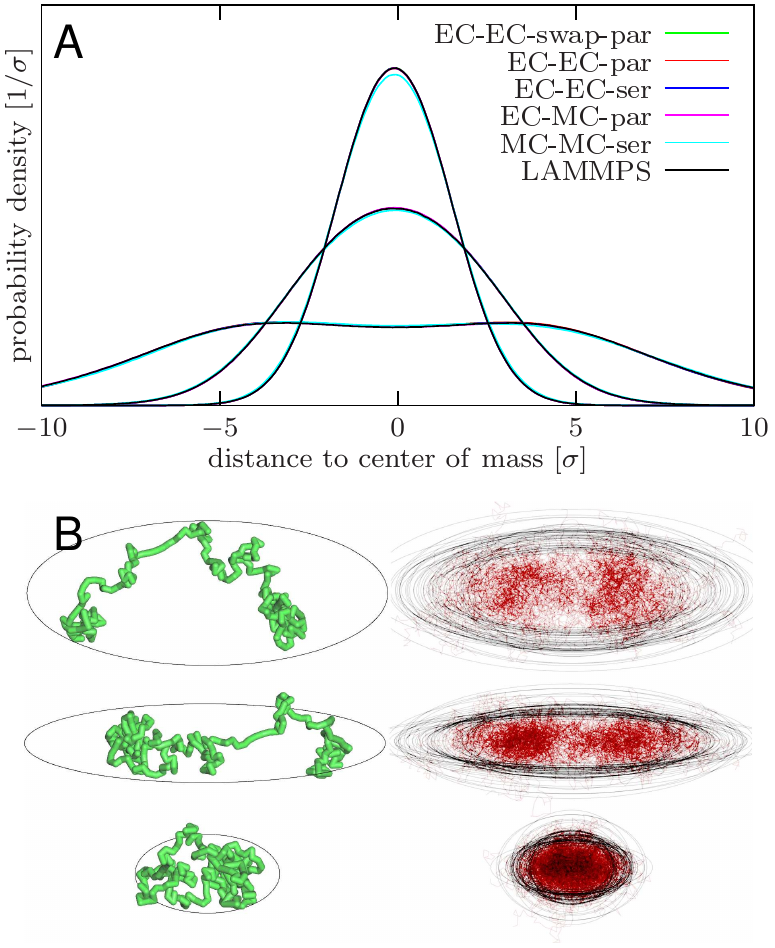}
 \caption{ 
(a) Bead distribution along the three principal axes of
   the moment of inertia tensor (for System I).  
(b) Snapshots of polymer configurations.
      The left column shows a single  configuration,  the right column
   an overlay of   50 configurations to visualize the distribution of the
     beads. Each row shows the projection to a plane spanned by two distinct
     principal axes. 
   The ellipsoids have the same
     moment of inertia tensor as the polymer.
}  
 \label{fig:traeg}
\end{figure}

The shape of a polymer in a dense melt is  ellipsoidal rather 
 than spherical, which can be shown by the distribution 
of beads with respect to the center of mass 
in the coordinate system which is given by the 
eigenvectors $\vec e_i$ of the moment of inertia tensor,
\begin{align}
 (\mat I)_{ij} = \sum_k (\vec r_k^2 \delta_{ij} - r_{k,i}r_{k,j}),
\end{align}
of a polymer.\cite{karayiannis2009}
The sum runs over all beads of a polymer, 
where $r_{k,i}$ denotes the $i$th component of the $k$th bead coordinate.

Following Ref.\ \citenum{karayiannis2009}, 
we can use the eigenvalues $I_1 \leq I_2 \leq I_3$ of the moment of inertia
tensor to characterize the shape in terms of its asphericity 
$b\equiv \frac{1}{2} (I_1+I_2)-I_3$, 
 acylindricity $c\equiv I_1-I_2$, and shape anisotropy 
$\kappa^2 \equiv  4(1-3(I_2I_3+I_3I_1+I_1I_2)/({\rm Tr}\,{\mat I})^2$.  
Additionally, there exist several analytical
predictions for an infinite freely jointed chain,\cite{solc1971}
\begin{align}
	\lim_{N \to \infty} 
 \frac{\langle 4b^2+3c^2 \rangle}{\langle {\rm Tr}\,{\mat I} \rangle^2 } 
   = \frac{2}{3},
\label{eq:solc}
\end{align}	
and for\cite{koyama1968}
\begin{align}
  \lim_{N \to \infty} \frac{\langle I_1+I_2-I_3 \rangle}
   {\langle {\rm Tr}\,{\mat I}\rangle} &= 0.754,
  \nonumber \\
 \lim_{N \to \infty} \frac{\langle I_1-I_2+I_3 \rangle}
   {\langle {\rm Tr}\,{\mat I}\rangle} &= 0.175, 
  \nonumber \\
	\lim_{N \to \infty} \frac{\langle -I_1+I_2+I_3 \rangle}
  {\langle {\rm Tr}\,{\mat I}\rangle} &= 0.0646,
\label{eq:koyama}
\end{align}
which can be tested.  

In Fig.\ \ref{fig:traeg}, we compare the distribution of
beads of one polymer in the system spanned by the 
eigenvectors of  the moment of inertia tensor for
all algorithms (for System I). 
  The  different widths of the distributions 
 along the three principal axes of
   the moment of inertia tensor in  Fig.\ \ref{fig:traeg}(a)
   implies that polymers in the melt 
    have an ellipsoidal shape. The
   distribution along the largest eigenvalue axis is  bimodal
  corresponding to an additional  dumbbell-like shape in this direction. 
The agreement between all simulation algorithms is excellent.
Our results also agree with MC simulation results 
 in Ref.\ \citenum{karayiannis2009}.
In Fig.\ \ref{fig:traeg}(b), we visualize the actual 
shapes of polymers  demonstrating the prolate shape of a polymer. 
   Snapshots in the first two rows 
   confirm the the  dumbbell-like shape with a 
  minimum in the bead distribution
     along the largest eigenvalue axis.

In Table \ref{tab:shape}, we compare the shape descriptors
from our Monte-Carlo schemes and LAMMPS with the theoretical expectations
(\ref{eq:solc}) and (\ref{eq:koyama}). All results coincide very well.

\begin{table*}
\begin{ruledtabular}
\begin{tabular}{lrrrrrrrrrr}
 & \multicolumn{3}{c}{MC-MC-ser} & \multicolumn{3}{c}{EC-EC-swap-par} & \multicolumn{3}{c}{LAMMPS} & theo.\\
 &  Sys.I & Sys.II & Sys.III &  Sys.I & Sys.II & Sys.III  &  Sys.I & Sys.II & Sys.III &  \\
\colrule 
$\langle \kappa^2 \rangle$ &  $0.390$ & $0.388$ & $0.387$ &  $0.399$ & $0.397$ & $0.395$  &  $0.402$ & $0.397$ & $0.395$  & $0.42$ \\
$\langle I_1+I_2-I_3 \rangle/\langle Tr(\mat I)\rangle$ &  $0.763$ & $0.758$ & $0.757$ &  $0.765$ & $0.764$ & $0.766$  &  $0.765$ & $0.767$ & $0.763$  & $0.754$ \\
$\langle I_1-I_2+I_3 \rangle/\langle Tr(\mat I)\rangle$ &  $0.173$ & $0.178$ & $0.177$ &  $0.172$ & $0.173$ & $0.172$  &  $0.172$ & $0.170$ & $0.173$ & $0.175$ \\
$\langle -I_1+I_2+I_3 \rangle/\langle Tr(\mat I)\rangle$ &  $0.0642$ & $0.06473$ & $0.06540$ &  $0.06284$ & $0.06310$ & $0.06251$  &  $0.06225$ & $0.06302$ & $0.06427$  & $0.0646$ \\
$\langle 4b^2+3c^2 \rangle/\langle Tr(\mat I) \rangle^2 $ &  $0.653$ & $0.621$ & $0.629$ &  $0.640$ & $0.638$ & $0.672$  &  $0.646$ & $0.643$ & $0.633$  & $0.667$ \\
\end{tabular}
\end{ruledtabular}
\caption{Comparison of  shape descriptors
from MC algorithms, LAMMPS, and  theoretical expectations.
For clarity, we only show the results from MC-MC-ser,
 EC-EC-swap-par and LAMMPS algorithms; 
  the other variants do not differ significantly.}
\label{tab:shape}
\end{table*}

\subsection{Distribution of the End-to-End Distance}

\begin{figure}[t]
 \includegraphics[width=0.95\linewidth]{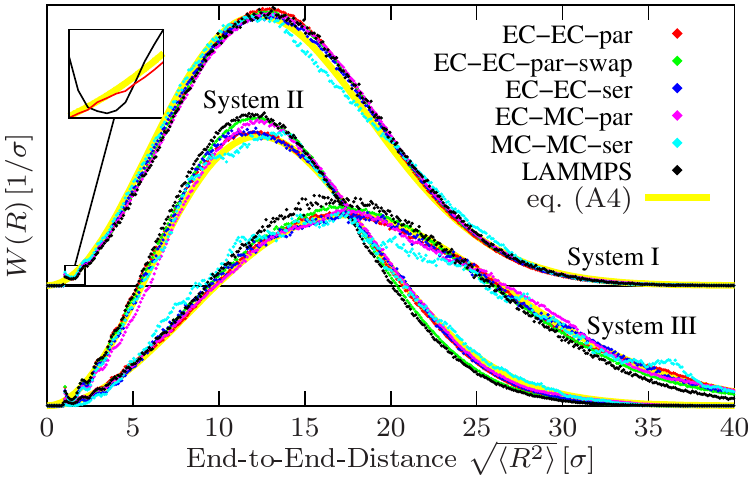}
 \caption{ 
  Distribution of the end-to-end distance $W_L(R)$ and 
 in comparison with eq.\ (\ref{eq:rquad_dist}) for ideal chains. 
 For  clarity the results for System I are shifted by an
   offset.   
 For short distances $W_L(R)$  exhibits  oscillations. 
  The inset shows 
  the ratio $W_L(R) /g(r)$ (red points) with 
 the  pair correlation $g(r)$, which shows no  oscillations.
 }
 \label{fig:rquad_dist}
\end{figure}

The distribution of the end-to-end distance $W(R)$ for an
ideal chain  with a 
mean-square end-to-end distance 
$\langle R^2 \rangle (N) = c N\sigma^2$ (see section \ref{sec:initial}
for the definition of the stiffness parameter $c$) 
is approximately given by a Gaussian  
distribution \cite{degennes,doiedwards}
\begin{align}
W(R)\mathrm{d}R =  4\pi R^2 \left( \frac{3}{2\pi cN\sigma^2}\right)^{3/2} 
   \exp\left( -\frac{3 R^2 }{2 cN\sigma^2}  \right) \mathrm{d}R.
 \label{eq:rquad_dist}
\end{align}
Therefore, the ideality of chains in a polymer melt 
can be tested by comparing simulation results for the 
 distribution of the end-to-end distance
with the Gaussian expectation (\ref{eq:rquad_dist}), see
 Fig.\ \ref{fig:rquad_dist}. 
The agreement with the Gaussian expectation is 
 indeed good, apart from an
oscillating behavior at small distances $R$. 
These oscillations can be explained by the 
influence of the pair correlation function $g(r)$ characterizing  the 
additional local liquid-like ordering of neighboring polymer beads. 
These oscillations are in quantitative agreement 
with  $g(r) W_L(R)\mathrm{d}R$, where we determined the 
pair correlation  $g(r)$  of beads in the polymer melt  numerically.

Also the agreement among the results for 
different  simulation algorithms in Fig.\ \ref{fig:rquad_dist} is very good.
Only the standard serial MC-MC algorithm shows deviations
because of its long equilibration times.

\end{document}